\documentclass[runningheads]{llncs}
\usepackage[T1]{fontenc}

\usepackage{url}
\usepackage{amsmath, amsfonts, amssymb, booktabs}
\usepackage{algorithm}
\usepackage{graphicx}
\usepackage{breqn}
\usepackage{listings}
\usepackage{algpseudocode}
\usepackage{multirow}
\usepackage{footnote}

\begin{document}
\mainmatter              
\title{A Social Data-Driven System for Identifying Estate-related Events and Topics}
\titlerunning{Identifying Estate-related Events and Topics on Social Media}  
%
\author{Wenchuan Mu\inst{1}\orcidID{0009-0007-2395-9731}
\and Menglin Li\inst{1}\orcidID{0000-0002-7890-7636}
\and Kwan Hui Lim\inst{1}\orcidID{0000-0002-4569-0901}}

\authorrunning{W. Mu et al.}  

\institute{
Singapore University of Technology and Design\\
\email{\{wenchuan\_mu, menglin\_li, kwanhui\_lim\}@sutd.edu.sg}}

\maketitle 

\begin{abstract}
Social media platforms such as Twitter and Facebook have become deeply embedded in our everyday life, offering a dynamic stream of localized news and personal experiences. The ubiquity of these platforms position them as valuable resources for identifying estate-related issues, especially in the context of growing urban populations. In this work, we present a language model-based system for the detection and classification of estate-related events from social media content. Our system employs a hierarchical classification framework to first filter relevant posts and then categorize them into actionable estate-related topics. Additionally, for posts lacking explicit geotags, we apply a transformer-based geolocation module to infer posting locations at the point-of-interest level. This integrated approach supports timely, data-driven insights for urban management, operational response and situational awareness.
\end{abstract}

\section{Introduction}

Over the past two decades, social media platforms such as Twitter/X and Facebook have undergone unprecedented growth, now reaching approximately 82\% of the global online population, with nearly 20\% of users’ online time spent on these platforms~\cite{comscore}. These platforms have evolved into essential channels for real-time information dissemination and discussion, encompassing topics ranging from entertainment and pop culture to more specialized domains such as politics and human rights. In parallel, organizations increasingly leverage social media as a sensing modality to identify and monitor both large-scale events (e.g., natural disasters, public crises) and localized issues (e.g., infrastructure faults, community disturbances).

Despite the value of social media as an information source, its scale and velocity introduce significant challenges, foremost among them being information overload~\cite{fu2020social}. This impairs the end users’ abilities to efficiently filter, retrieve, and contextualize relevant content. In response, a range of social media analytics systems have emerged. Examples include InfoTrace~\cite{Cheng-JCDL24}, which tracks the lifecycle of social media campaigns; DISCO~\cite{fu2022disco}, a framework for explainable disinformation detection; RAPID~\cite{lim-ecml18}, which enables real-time mining of streaming social media data; Li et al.~\cite{li2017real} that presented a framework implementing clustering and temporal identification for event detection; and Rosa et al.~\cite{rosa2020event} that utilizes user behaviour changes over time to detect pandemic-related events on social media. While these systems address important facets of social media analysis, a critical gap remains: the automated detection of estate-related events, such as facility breakdowns, noise complaints, or parking violations, within both historical and real-time social media data streams.

To address this gap, we propose a novel system for detecting estate-related events and associated discussion topics from both archival and live social media data. This system is part of the broader Estate-IQ initiative, which aims to automate estate operations and maintenance through AI-driven event detection and decision support.

\begin{figure*}[t]
    \centering
    \includegraphics[width=\linewidth]{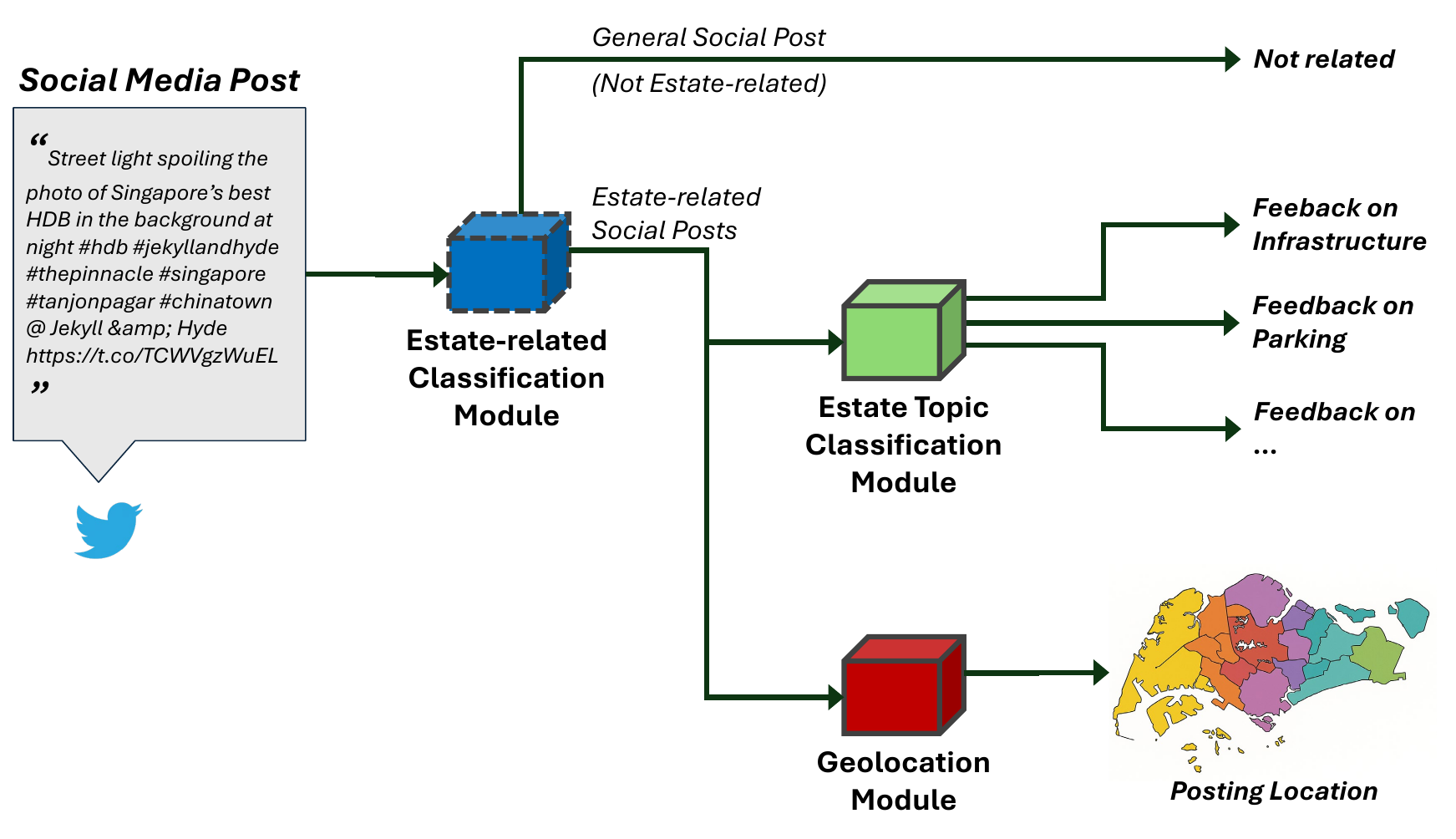}
    \caption{Model Architecture of Our Proposed System}
    \label{fig:model}
\end{figure*}

\section{Problem Definition}
\label{problemDef}

The ubiquity of smartphones and mobile connectivity has transformed social media platforms into pervasive channels for both real-time news dissemination and the sharing of daily personal experiences. However, the high velocity and volume of user-generated content create significant challenges in filtering and identifying estate-related information amidst the background noise of general posts.

To address these challenges, we formally define two key tasks: \textit{Estate-related Post Detection} and \textit{Estate Topic Classification}. Let $S = \{s_1, \ldots, s_n\}$ denote a collection or stream of $n$ social media posts, where $s_n$ represents the most recent entry. Each post $s_i$ is modeled as a sequence of $T$ tokens:
\[
s_i = \{s_i^1, \ldots, s_i^T\}
\]

\textbf{Estate-related Post Detection.} We define the estate detection task as a binary classification problem:
\[
D_E = (S, L^E)
\]
where $L^E = \{L_1^E, \ldots, L_n^E\}$ is the set of binary labels with $L_i^E \in \{0,1\}$. A label of $1$ denotes that $s_i$ contains estate-related content (e.g., facility faults, noise complaints), and $0$ indicates otherwise. The objective is to learn a classifier:
\[
C_D^E : S \rightarrow L^E
\]
that accurately maps each social media post to its corresponding estate relevance label.

\textbf{Estate Topic Classification.} For posts identified as estate-related, we define a secondary multi-class classification task:
\[
D_T = (S, L^T)
\]
where $L^T = \{L_1^T, \ldots, L_n^T\}$ and $L_i^T \in \{0,1,2,3\}$. Each label corresponds to a specific estate-related topic: Infrastructure, Parking, Noise and Others.

The corresponding classifier is defined as:
\[
C_D^T : S \rightarrow L^T
\]
which assigns an appropriate estate topic to each post previously identified as relevant by $C_D^E$.

The overarching goal is to build a pipeline that first filters estate-related content from the general stream of social media data and then categorizes the filtered posts into actionable topic domains. Together, these models form the backbone of an automated estate event detection system, enabling real-time urban situational awareness.

\section{System Architecture}
Our proposed system is composed of four core components: (i) \textbf{Data Repository/Stream}, (ii) \textbf{Estate-related Post Classification}, (iii) \textbf{Estate Topic Classification} and (iv) \textbf{Social Post Geolocation}. Figure~\ref{fig:model} provides an overview of the system architecture and its data flow. In the subsections below, we elaborate on each component in greater detail.

\subsection{Data Repository/Stream}

As introduced in Section~\ref{problemDef}, our framework operates over a collection of social media posts, which may originate from either a static repository or a live data stream. Each input to the system consists of a single post in textual form, drawn from this larger collection. The system is designed to process posts in real-time or in batch mode, supporting both retrospective and online analyses. For the purpose of applying this work, personal identifiers from the datasets are anonymized to an anonymous ID that still enables us to identify posts by the same user but is otherwise not mappable to the real-life user.

\subsection{Estate-related Post Classification}

To detect estate-relevant content within social media posts, we utilize the Bidirectional Encoder Representations from Transformers (BERT) model~\cite{devlin-etal-2019-bert}. BERT is a multi-layer bidirectional Transformer encoder trained using self-supervised objectives such as Masked Language Modeling (MLM) and Next Sentence Prediction (NSP). 

For our binary classification task, we fine-tune BERT on a curated dataset of annotated social media posts labeled as estate-related or not. The output of this component is a binary prediction indicating whether a given post pertains to estate-related content. Posts identified as relevant are then passed to the next component for fine-grained topic categorization. Experimental results in the later sections will justify our choice of BERT over other baselines methods.

\subsection{Estate Topic Classification}

Posts classified as estate-related undergo further classification into specific topical categories. This component also uses the BERT model, fine-tuned on a proprietary dataset comprising maintenance reports submitted by estate residents~\cite{Mu-IJCNN24}. While the original dataset contains 13 distinct categories, empirical analysis shows that over 94.8\% of reports fall into three major categories: \textit{Infrastructure}, \textit{Parking}, and \textit{Noise}. 

Accordingly, we consolidate the remaining 10 less frequent categories into a single class labeled \textit{Others}, resulting in a four-class classification task. This setup enables a balance between model simplicity and topic coverage, ensuring robust classification across the most prevalent estate-related concerns.

\subsection{Social Post Geolocation}
For estate-related posts lacking explicit geotags, we apply an additional geolocation module to infer the likely posting location. We adopt the transformer-based framework, transTagger, proposed in our prior work~\cite{li2023transformer}, which builds upon pre-trained language models and integrates both textual and non-textual cues for fine-grained location inference at the POI level. The framework categorizes inputs by data modality, applies an optimal feature fusion strategy, and incorporates a hierarchical temporal encoding scheme. A concatenated representation of positional embeddings is used to better capture fine-grained spatiotemporal context. This component enables spatial contextualization of posts, supporting downstream applications like estate monitoring and geospatial event mapping.

\section{Experimental Results}

We conducted a preliminary evaluation of the core components of our system using a Twitter/X dataset and a proprietary estate incident dataset, both collected from the same geographical region.

Table~\ref{tabEstateDetect} presents the performance of various models for detecting estate-related posts. We evaluated five methods: Logistic Regression (LogReg), Recurrent Neural Networks with GloVe embeddings (RNN+GloVe)\cite{pennington2014glove}, RNN with Word2Vec embeddings (RNN+W2V)\cite{mikolov2013distributed}, BERT, and MPNet~\cite{song2020mpnet}. Among these, BERT achieved the highest accuracy and F1-score, establishing it as the most effective architecture for this task. As such, we adopt BERT as the classification backbone in our deployed system.

\begin{table}[htbp]
\begin{minipage}{.53\linewidth}
\centering
\caption{Estate-related Post Classification}
\begin{tabular}{lcc}
\hline
\textbf{Model} & \textbf{Accuracy} & \textbf{F1-score} \\ \hline
LogReg          & 0.810         & 0.300        \\ 
RNN+GloVe        & 0.500         & 0.375       \\ 
RNN+W2V         & 0.300         & 0.462       \\ 
BERT            & 0.950         & 0.800        \\ 
MPNet           & 0.850         & 0.595      \\ \hline
\end{tabular}
\label{tabEstateDetect}
\end{minipage}%
\begin{minipage}{.47\linewidth}
\centering
\caption{Estate Topic Classification}
\begin{tabular}{lcc}
\hline
\textbf{Estate Topic} & \textbf{Accuracy} & \textbf{F1-score} \\ \hline
Infrastructure & 0.940 & 0.877 \\
Parking               & 0.985 & 0.971 \\
Noise                         & 0.837 & 0.829 \\
Others                        & 0.188 & 0.297 \\  
\textbf{Weighted Avg}     & \textbf{0.882} & \textbf{0.865} \\ \hline
\end{tabular}
\label{tabTopicDetect}
\end{minipage}%
\end{table}

For the downstream task of estate topic classification, we employed BERT and evaluated performance across four topical categories: Infrastructure, Parking, Noise, and Others. As shown in Table~\ref{tabTopicDetect}, BERT yielded strong performance for the first three categories, with both Accuracy and F1-score exceeding 80\%. Although performance for the Others category was comparatively lower, this class constitutes only 5.2\% of the dataset and contributes minimally to the overall performance, as reflected in the weighted average scores.

We also evaluated the geolocation module using results from our previous work, transTagger~\cite{li2023transformer}. On the Singapore Twitter/X dataset, transTagger achieved an accuracy of 0.691 and a mean geotagging distance error of 2.21 km. While the accuracy may appear moderate, the geolocation task is inherently challenging due to the presence of 9,666 distinct Points-of-Interest (POIs). To mitigate privacy risks, we employ a variant of transTagger that performs geotagging at the neighbourhood granularity rather than at the POI level.

\section{Conclusion and Future Work}

In this paper, we presented a social data-driven platform that leverages a pre-trained language model to perform hierarchical classification of social media content. Our system first detects estate-related posts and subsequently categorizes them into specific topics of interest. This framework is designed to address the growing challenge of information overload on social media by automatically surfacing relevant, actionable content for estate management and urban operations. By prioritizing such posts, the system facilitates timely interventions and supports more efficient decision-making in densely populated environments. As the system relies on geotagged data and geolocation models, future work will focus on systematically examining the implications of privacy risks and representational bias in both classification and geospatial inference tasks~\cite{yang2020protecting,malik2015population}.

\vspace{5mm}
{\small
\noindent\textbf{Acknowledgment.}
This research/project is supported by the National Research Foundation (NRF), Singapore, and Ministry of National Development (MND), Singapore under its Cities of Tomorrow R\&D Programme (CoT Award COT-V2-2020-1). Any opinions, findings and conclusions or recommendations expressed in this material are those of the author(s) and do not reflect the views of NRF and MND. 
K. H. Lim is also supported by a MOE AcRF Tier 2 (MOE-T2EP20123-0015).
The authors would also like to thank Xjera Labs for collaborating on this project.
}

%
%
\bibliographystyle{splncs04}
\bibliography{sample-base}

\end{document}